# High-speed and high-responsivity hybrid silicon/black-phosphorus waveguide photodetectors at 2 μm

Yanlong Yin[1], Rui Cao[2], Jingshu Guo[1], Chaoyue Liu[1], Jiang Li[1], Xianglian Feng[1], Huide Wang[2], Wei Du[1], Akeel Qadir[3], Han Zhang[2], Yungui Ma[1], Shiming Gao[1], Yang Xu[3], Yaocheng Shi[1], Limin Tong[1], and Daoxin Dai[1,†]

1. State Key Laboratory for Modern Optical Instrumentation, College of Optical Science and Engineering, Zhejiang University, Zijingang Campus, Hangzhou, China.

2. Shenzhen Engineering Laboratory of Phosphorene and Optoelectronics Collaborative Innovation Center for Optoelectronic Science and Technology and Key Laboratory of Optoelectronic Devices and Systems of Ministry of Education and Guangdong Province Shenzhen University Shenzhen 518060, P. R. China.

3. College of Information Science and Electronic Engineering, Zhejiang University, Hangzhou, Zhejiang, 310027 China.

Correspondence and requests for materials should be addressed to D. D. (dxdai@zju.edu.cn)

## Abstract

Silicon photonics is being extended from the near-infrared (near-IR) window of 1.3-1.5 μm for optical fiber communications to the mid-infrared (mid-IR) wavelength-band of 2 μm or longer for satisfying the increasing demands in many applications. Mid-IR waveguide photodetectors on silicon have attracted intensive attention as one of the indispensable elements for various photonic systems. Previously high-performance waveguide photodetectors on silicon were realized for the near-IR window of 1.3-1.5 μm by introducing another semiconductor material (e.g., Ge, and III-V compounds) in the active region. Unfortunately, these traditional semiconductor materials do not work well for the wavelength of ~2 μm or longer because the light absorption becomes very weak. As an alternative, two-dimensional materials provide a new and promising option for enabling active photonic devices on silicon. Here black-phosphorus (BP) thin films with optimized medium thicknesses (~40 nm) are introduced as the active material for light absorption and silicon/BP hybrid *ridge* waveguide photodetectors are demonstrated with a high responsivity at a low bias voltage. And up to 4.0Gbps data transmission is achieved at 2μm.

**Introduction**

In the past decade silicon photonics has been very promising because of its unique advantages[1], e.g., the CMOS compatibility, the ultra-high integration density, etc. Various silicon photonic integrated devices have been developed successfully for the applications of optical fiber communications operating with the near-infrared (near-IR) wavelength-band of 1.31/1.55 μm[2,3]. More recently the mid-infrared (mid-IR) wavelength-band of 2-20 μm[4] has also been becoming very attractive for many important applications in optical communication[5], nonlinear photonics[6], lidar[7], and optical bio-sensing[8]. For example, for some bio-molecules and gases such as glucose, $CH_4$, $CO_2$ and CO, there are strong overtone and combination absorption lines in the wavelength range of 2-3μm[9]. In order to satisfy these increasing demands, silicon photonics is being extended to the mid-IR range including the 2 μm wavelength-band. Some low-loss mid-IR silicon photonic waveguides have been proposed[10-15], which were utilized further for realizing some mid-IR silicon photonic devices and circuits, e.g., nonlinear photonic chips[16], and arrayed- waveguide gratings[17,18], etc.

However, it is still a big challenge to realize high- performance waveguide-type photodetectors on silicon for the wavelength-band of 2 μm or beyond. As it is well known, most traditional semiconductor materials, which work very well at 1.3/1.55 μm[19], cannot be extended for longer wavelength because their cut-off wavelength for light absorption is limited by the bandgap and cannot be modified freely. In order to realize the photodetection of longer wavelength, a hybrid silicon/GaInAsSb photodetector was demonstrated with a responsivity of 0.44 A/W at 2.29 μm by using the adhesive bonding process with the DVS-BCB (Benzocyclobutene) as an adhesive bonding agent[20,21]. In this device, the coupling between the silicon waveguide and the photodiode waveguide strongly depends on the thicknesses of the BCB layer and the intrinsic absorbing region, and thus one needs to control the phase matching by carefully optimizing the layer thicknesses[19]. More recently, the defect-level absorption in silicon was utilized to realize a monolithic silicon photodiode with a high speed of 15 GHz for the 2 μm wavelength-band[22]. One should notice that a bias voltage of as high as ~30 V is needed in this case and it is also hard to be extended further for longer wavelengths.

Two-dimensional (2D) materials have attracted intensive attention owning to their unique properties, such as ultra-high carrier mobility and strong light absorption in a broad band[23-25]. Therefore, in recent years hybrid silicon/ 2D-material waveguide photodetectors have become very attractive for infrared photodetection[26-39]. Currently most of them were reported for the 1.55 μm wavelength-band, while the responsivity is 30-100 mA/W and the 3dB-bandwidth is 18-70 GHz [26-31] (see Table S1 in Supplementary). With the help of hexagonal boron nitride and metal-graphene edge contact, an excellent hybrid silicon/graphene waveguide photodetector at 1.55 μm was demonstrated with a high responsivity of 0.36 A/W and a high 3 dB-bandwidth of 42 GHz[31]. Black-phosphorus (BP) is also becoming attractive for the infrared photodetection on silicon[32-33]. For example, recently a silicon/BP waveguide photodetector at 1.55 μm was realized with a responsivity of 0.135 A/W and a 3dB-bandwidth of ~3 GHz[32]. Fortunately, graphene has a zero bandgap[25] and BP has a small bandgap of 0.3-1 eV[34,35]. Therefore, it is promising for realizing the photodetection for the wavelengths of 2 μm and beyond (see Table S1 in Supplementary). Previously a hybrid silicon/graphene waveguide photodetector at 2.75 μm was demonstrated with 0.6 A/W at −1.5V bias voltage[36]. However, high-speed photodetection has not been reported yet. More recently, several

*normal-incident* BP photodetectors working at 2-3.9 μm were reported, while the responsivity and the speed are still low[37-39]. Currently no results have been reported for silicon/BP waveguide photodetectors at the wavelength of 2 μm or longer, which is a key in mid-IR silicon photonics for many applications[40]. In this paper, we propose and demonstrate hybrid silicon/BP waveguide photodetectors at 2 μm experimentally. Here silicon-on-insulator (SOI) ridge waveguides are used and the BP thin films with a medium thickness of ~40 nm are chosen optimally for achieving sufficient light absorption, low mode mismatching loss, high mechanical reliability, as well as easy BP transfer process. In particular, an imprint-transfer process of BP was developed so that the transferred BP thin film can cover the top-surface and the sidewalls of the SOI ridge waveguide well, which enables the light absorption enhancement for the fundamental mode of TE polarization (TE$_0$) in the SOI ridge waveguide. For the present silicon/BP waveguide photodetectors, the measured responsivity is as high as ~306.7 mA/W at a low bias voltage of 0.4 V, and the 3 dB-bandwidth is as high as 1.33 GHz. Finally, a 4.0 Gbit/s data-receiving test is also demonstrated *experimentally*.

## Structure of Si/BP hybrid waveguide photodetectors

Fig. 1(a) shows the schematic configuration of the present hybrid silicon/BP waveguide photodetectors at 2 μm. In this design, an SOI ridge waveguide is adopted. The silicon-ridge height and the silicon-core height were chosen as 220nm and 340nm, respectively. A 20nm-thick Al$_2$O$_3$ layer was deposited by the atom layer deposition (ALD) method to cover the SOI ridge waveguide as the insulator between the BP and the silicon core. And two Ti/Au (15 nm/80 nm) metal pads were symmetrically located with a 3.6 μm separation at both sides of the silicon ridge as the drain and the source electrodes [see Fig. 1(a)]. The BP thin film was then transferred on the top of the SOI ridge waveguide to cover the silicon ridge as well as the electrodes by using the imprint-transfer process (see the details in *Method* and the Supplementary section S2). Finally, in order to prevent the BP thin film from oxidization, a 500nm-thick PMMA upper-cladding was spin-coated on the chip. Fig. 1(b) shows the microscopy images of three samples ($S_1$, $S_2$, and $S_3$), for which the lengths of the waveguides covered by the BP thin film are about 35 μm ($S_1$), 36.7 μm ($S_2$) and 33.3 μm ($S_3$), respectively. The BP thicknesses are respectively about 36.8nm, 47.5nm, and 48.5nm, which were measured with an atomic force microscopy (AFM), as shown in Fig. 1(c)-(e).

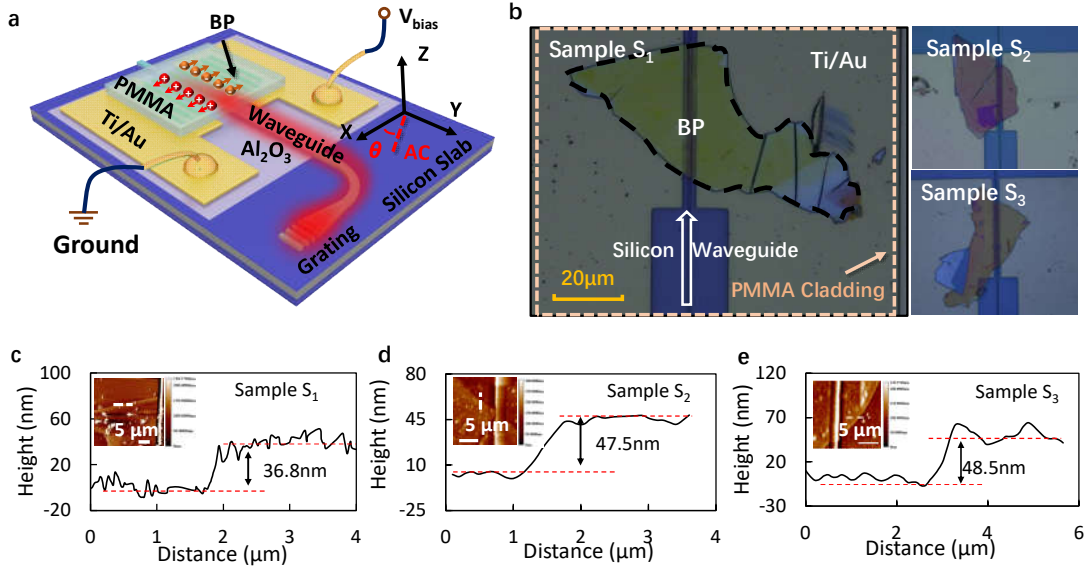

**Figure 1 | Hybrid silicon/BP waveguide photodetectors. a,** Schematic configuration of the present hybrid silicon/BP waveguide photodetector at 2 μm; **b,** Microscopy images for three samples (S$_1$, S$_2$, and S$_3$). The lengths of the waveguides covered by the BP films are 35 μm (S$_1$), 36.7 μm (S$_2$) and 33.3 μm (S$_3$), respectively; **c-e,** The thickness measurement results by AFM for these three samples. The BP thicknesses $h_{BP}$ are 36.8 nm, 47.5 nm, and 48.5 nm, respectively. **Inset,** the AFM images and scanning length denoted by the while dotted line.

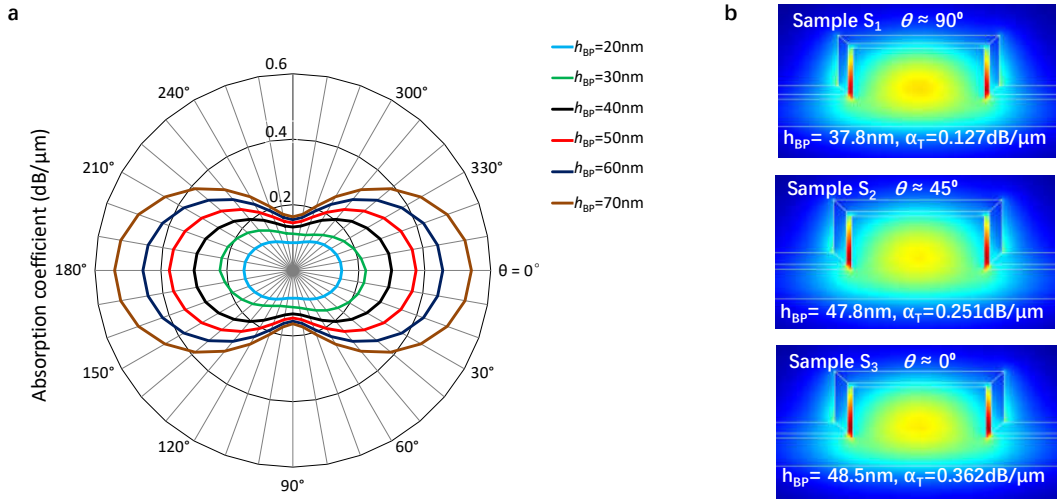

**Figure 2 | Optical absorption properties of the hybrid silicon/BP waveguide Photodetectors. a,** Calculated light absorption coefficient $\alpha$ as the BP orientation angle $\theta$ varies from 0° to 360°. Here the BP thickness is chosen as $h_{BP}$=20, 30, 40, 50, 60, and 70nm; **b,** Calculated electric field profiles of the TE$_0$ mode supported in the hybrid silicon/BP waveguides and absorption loss $\alpha_T$ for samples S$_1$, S$_2$ and S$_3$ at 2μm.

**BP films characterization and mode analysis**
Since the BP material is anisotropic[41], the thickness and the orientation of the BP thin film play important roles on the electronic and optical properties of the photodetectors. For the present hybrid silicon/BP waveguide photodetectors, we choose the $TE_0$ mode, for which the electric component $E_x$ is dominant. Fig. 2(a) shows the calculated light absorption coefficient α as the BP orientation angle $\theta$ varies from 0° to 360°, where $\theta$ is defined as the angle between the armchair axis and the x direction (see the coordinate system shown in Fig. 1(a) and the details for the theoretical model in Section

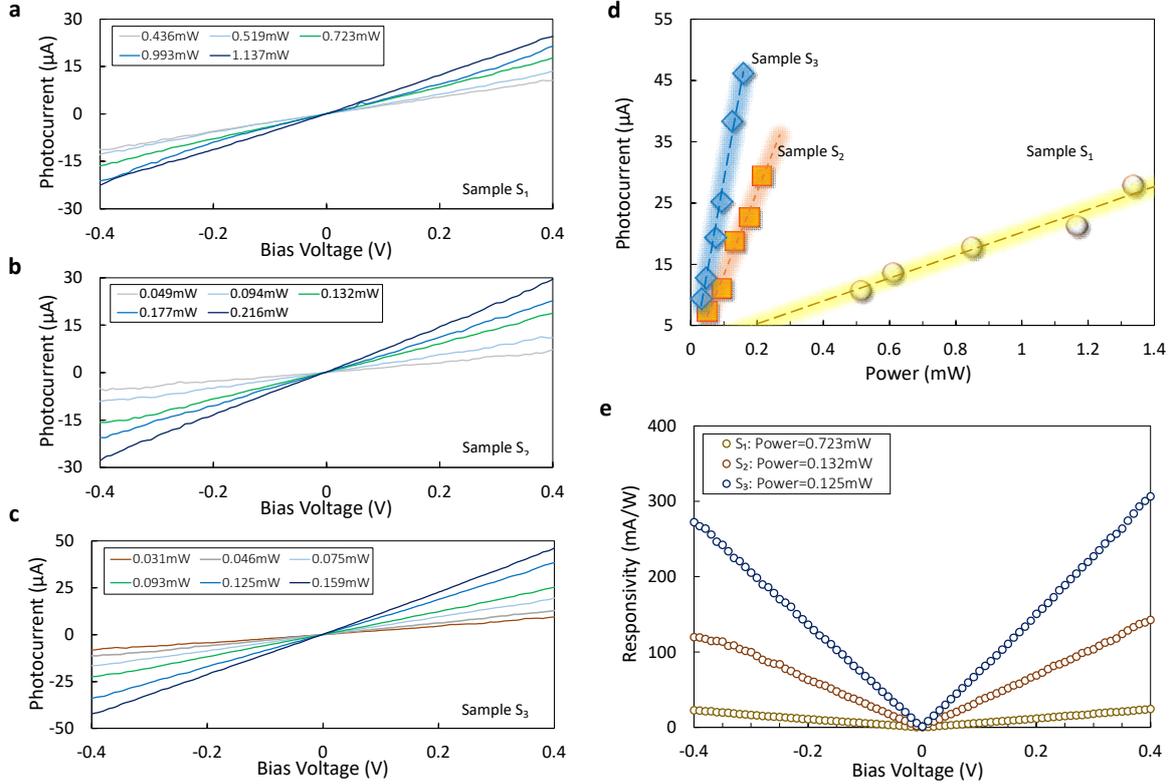

**Figure 3| Static responses of hybrid silicon/BP waveguide photodetectors. a-c,** Measured IV curves for the silicon/BP waveguide photodetectors of $S_1$ (a), $S_2$ (b) and $S_3$ (c) as the bias voltage varies from −0.4V to 0.4V; **d,** Measured photocurrents for samples $S_1$, $S_2$ and $S_3$ operating at $V_{bias}$=0.4 V as the optical power increases. **e,** Responsivities for $S_1$, $S_2$ and $S_3$ as the bias voltage varies.

4 of Supplementary). Here the BP thickness is chosen as $h_{BP}$ = 20, 30, 40, 50, 60, and 70 nm for the calculation. It can be seen that the absorption coefficient α in the case of $\theta$=0° (or 180°) increases from 0.146 dB/μm to 0.574 dB/μm as the BP thickness $h_{BP}$ increases from 20nm to 70nm. In contrast, the absorption coefficient α becomes much smaller when $\theta$=90° (or 270°). For this case, the absorption coefficient α varies from 0.084 dB/μm to 0.179 dB/μm as the BP thickness $h_{BP}$ increases from 20 nm to 70 nm. Apparently, one can achieve stronger light absorption by choosing a thicker BP thin film, and consequently the absorption length for the waveguide photodetector can be minimized to enable high bandwidth. On the other hand, a thin BP film is preferred to avoid high mode-mismatching loss and reflection loss at the junction between the passive section to the active section of the SOI ridge waveguide. In this paper we choose a medium BP thickness of ~40 nm for the hybrid silicon/BP waveguide photodetectors. Such a BP thin film is also preferred because it has higher mechanical reliability than ultra-thin BP films (e.g., $h_{BP}$ <10 nm), so that it can easily be transferred to cover the SOI ridge waveguide. Furthermore, this kind of ~40 nm-thick BP thin film also has high flexibility, and it is even possible to cover the top-surface and the sidewalls of the SOI ridge waveguide very well when transferred by using the imprint-transfer technique. In this way, the light absorption of the TE polarization mode in the SOI ridge waveguide can be enhanced significantly. In contrast, it is usually difficult to cover the sidewalls for ~100nm-thick BP films due to the inflexibility. In our experiment, the BP thicknesses for the three fabricated photodetectors are respectively about 36.8nm, 47.5nm, and 48.5nm. The orientation angles for the BP films of samples $S_1$, $S_2$ and $S_3$ are around $\theta$=0°, 45°, and 90°, which were determined by the measured Raman spectrum (see Section 3 in Supplementary). Fig. 2(b) show the calculated electric-field profiles of the $TE_0$ mode supported in the hybrid silicon-BP waveguide for sample $S_1$, $S_2$, and $S_3$. It can be seen that they have different light absorption coefficients, i.e., $\alpha_{S1}$=0.127 dB/μm, $\alpha_{S2}$=0.251 dB/μm, and $\alpha_{S3}$=0.362 dB/μm. The light absorption coefficient was also characterized by measuring the transmissions at the rear of the hybrid silicon/BP waveguides with and without BP (see Section 5 in Supplementary). Here the metal absorption is negligible since the gap between the Ti/Au pad and the silicon ridge is as large as 1.8 μm for these photodetectors. The measured light absorption coefficients are $\alpha_{S1}$=0.105 dB/μm and $\alpha_{S3}$=0.502 dB/μm for samples $S_1$ and $S_3$, respectively, which agree well with the theoretical prediction.

**Static optoelectronic response of Si/BP hybrid waveguide photodetectors and temperature stability**
The static properties of the present hybrid silicon/BP photodetectors were characterized by using a setup with a 1.957 μm laser. Fig. 3(a)-(c) shows the measured IV curves of the three samples (i.e., $S_1$, $S_2$ and $S_3$) for the cases with different optical powers as the bias voltage $V_{bias}$ varies from −0.4 V to 0.4 V. It can be seen that the photocurrent increases linearly as the bias voltage $V_{bias}$ increases. Fig. 3(d) shows the measured photocurrents for samples $S_1$, $S_2$ and $S_3$ as the optical power increases. Here the bias voltage is fixed as $V_{bias}$=0.4 V. It can be seen that the photocurrent varies linearly as the optical power varies and the responsivities $R$ for the three samples are 21.4-26.2 mA/W, 116.2-146.9 mA/W, 260-306.7 mA/W, respectively. It can be seen that the responsivity $R$ is not sensitive to the optical power, which indicates that the present hybrid silicon/BP waveguide photodetector operates dominantly with the photovoltaic effect[32]. Otherwise, the responsivity for those photodetectors operating with the bolometric or photo conductive effect is strongly dependent on the input optical power[38]. Fig. 3(e) shows the responsivities for samples $S_1$, $S_2$ and $S_3$ as the bias voltage varies from −0.4 V to 0.4 V. It can be seen that the responsivity increases linearly as the bias voltage increases. It is expected to achieve a responsivity of higher than 1.5 A/W for sample $S_3$ when the bias voltage $V_{bias}$ is increased to 2.0 V. We also calculated the internal quantum efficiency (IQE) for these three photodetectors operating at $V_{bias}$ = 0.4 V as the optical power varies from 0.03 mW to 0.41 mW. Here the IQE is given by $\eta_i = Rhc/(\alpha_T \cdot e\lambda)$, where $\alpha_T$ is the BP absorption, $h$ is Planck constant, $c$ is the speed of light, $e$ is the electron charge, and $\lambda$ is the wavelength. It can be seen that samples $S_1$, $S_2$, and $S_3$ have

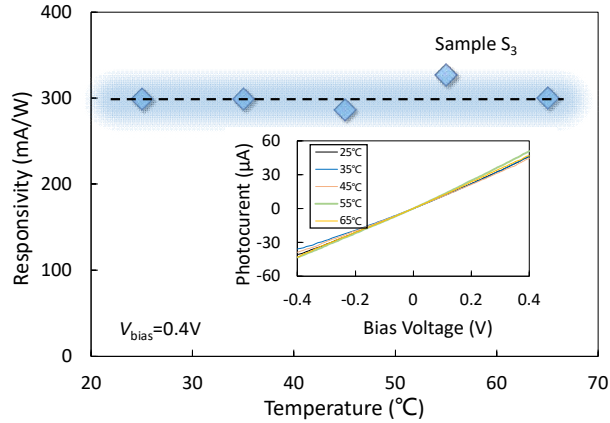

**Figure 4 | Temperature stability of hybrid silicon/BP waveguide photodetectors.** **The blue rhombus,** calculated responsivity as the temperature varies from 25 °C to 65 °C on the heating surface of Thermo Electric Cooler (TEC). **Inset,** Measured IV curves for sample $S_3$ under different stage temperature, 25 °C, 35 °C, 45 °C, 55 °C, 65 °C. Here the input optical power is 0.16 mW while the bias voltage $V_{bias}$ varies from −0.4 V to 0.4 V.

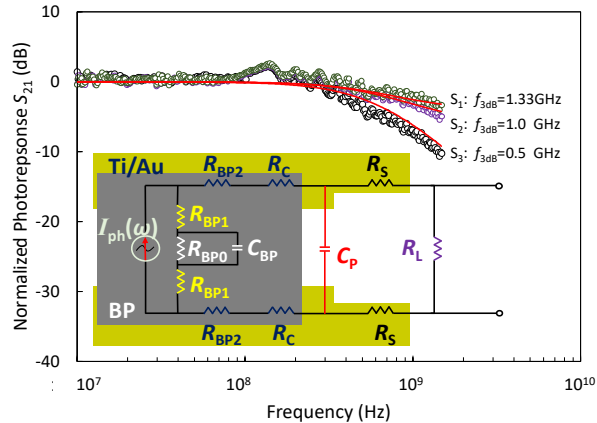

**Figure 5 | Measured frequency responses and the equivalent circuit model of hybrid silicon/BP waveguide photodetectors.** Measured results for the normalized frequency responses of the three samples when operating at $V_{bias}$ = 2.0 V. **Inset,** Established equivalent circuit model for the present photodetectors. **Red line,** Frequency responses calculated by using the equivalent circuit model with the fitted parameters (red line).

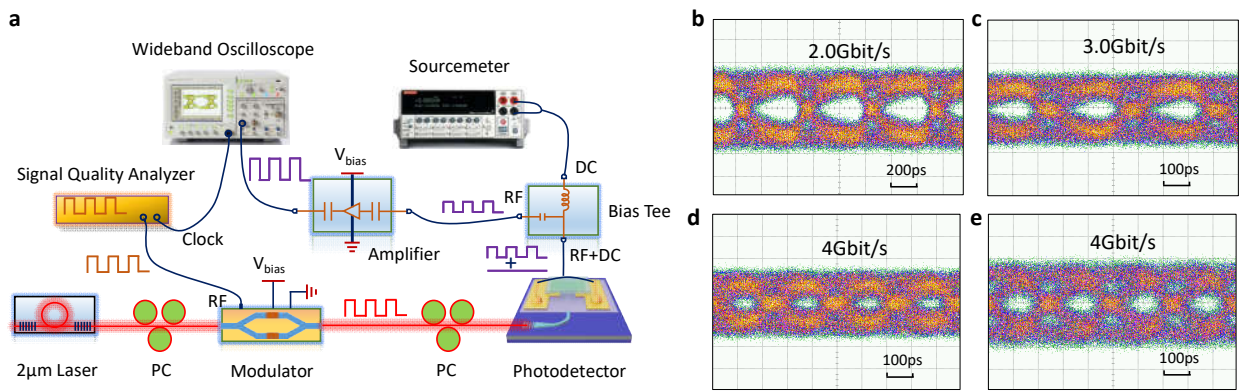

**Figure 6 | Measurement of eye-diagrams for the hybrid BP waveguide photodetector (Sample $S_2$). a,** Experimental setup for measuring eye-diagrams of the present hybrid silicon/BP waveguide photodetectors. **b-e,** Measured eye-diagrams with different bit rates. (b) 2.0 Gbit/s at 2 V; (c) 3.0 Gbit/s at

2 V; (d) 4.0 Gbit/s at 2 V; (e) 4.0 Gbit/s at 2.5 V.

different IQEs, which are 2.4-2.9%, 8.3-10.5%, and 16.6-20%, respectively. This is mainly due to the difference of the carrier recombination process in the BP films with different orientation angles and thicknesses, which verifies that the electronic and optical properties of the photodetectors strongly depend on the thickness and the orientation of the BP thin film.

In order to characterize the temperature stability of the fabricated photodetectors, the IV curves were measured for sample $S_3$ as an example when the stage temperature increases from 25 °C to 65 °C, as shown in Fig. 4. In this experiment, the input optical power is $P$=0.16 mW while the bias voltage $V_{bias}$ varies from −0.4 V to 0.4 V (see the inset). It can be seen that the photocurrent is not sensitive to the stage temperature. Correspondingly, the responsivity $R$ is kept constant almost and one has $R$=~300 mA/W even when the temperature increases to 65 °C, shown in Fig. 4(b). This makes it useful in potential for high-temperature operation.

## High speed photoresponse and electric circuit model

For the present hybrid silicon/BP waveguide photodetectors, the frequency responses were also characterized. In the experiment, a fiber laser at ~1.95 μm was used as the CW light source, which was modulated by using a high-speed LiNbO$_3$ optical modulator. Fig. 5 shows the measured results for the normalized frequency responses of these three samples ($S_1$, $S_2$ and $S_3$) when they were operating at $V_{bias}$ = 2.0 V. It can be seen that the 3dB-bandwidths for these three photodetectors are about 1.33, 1.0 and 0.5 GHz, respectively. For the present hybrid silicon/BP waveguide photodetectors, the total response time $\tau$ is estimated by[41],

$$\frac{1}{\tau} = \frac{1}{\tau_s + \tau_{tr}} + \frac{1}{\tau_r},$$

where $\tau_s$ is the saturation time, $\tau_{tr}$ is the transit time, and $\tau_r$ is the intrinsic recombination time. The transit time $\tau_{tr}$ is given as $\tau_{tr} = l^2/(2\mu V_{bias})$, where $l$ is the BP flake channel length, $\mu$ is the carrier mobility. According to this equation, it is expected to observe some strong dependence of the response time on the bias voltage if the transit time $\tau_{tr}$ is the dominant limitations for the response time. In our experiment, the measured frequency responses at different bias voltages (see Fig. S6 in Supplementary) are almost the same, which indicates that the carrier mobility is not the dominant factor limiting the response speed of the present photodetectors.

In order to understand the RC-constant limit, we established an equivalent circuit model for the present BP photodetector, as shown by the inset in Fig. 5 (see the details in Section 7 in Supplementary). With the measured RF reflection coefficient $S_{11}$, one can obtain the total impedance $Z_{in}$ of the photodetector according to the equation of $Z_{in}=Z_0(1+S_{11})/(1−S_{11})$, where $Z_0$=50 Ω. From the impedances $Z_{in}$ for samples $S_1$, $S_2$ and $S_3$ [see Fig. S7(a)-(d)], one can easily extract all the parameters ($R_C$, $R_S$, $R_{BP0}$, $R_{BP1}$, $R_{BP2}$, $C_P$, $C_{BP0}$ and $C_{BP1}$) for all the elements in the equivalent circuit (see Table S2 in Supplementary) by using the fitting technique based on genetic algorithms. With the equivalent circuit model with the fitted parameters, the frequency responses for the photodetectors were calculated in the frequency range from 10 MHz to 1.5 GHz, as shown by the red lines in Fig. 5. It can be seen that the calculation results agree very well with the measurement results. It is verified that the RC-constant is the limitation for the 3dB-bandwidth in the present case. It is possible to further improve the response speed by reducing the BP-metal contact resistance $R_c$, as well as the BP capacitance $C_{BP}$.

In our experiment the present hybrid silicon/BP waveguide photodetector was further used to receive high bit-rate data at 2 μm with the setup shown in Fig. 6(a). The signal from the pattern generator is transmitted into the RF port of the high-speed LiNbO$_3$ optical modulator. Fig. 6(b)-(e) show the measured eye-diagrams for the photodetector $S_2$ operating at $V_{bias}$=2 V. It can be seen that the eye-diagram is open even with a bit rate as high as 3.0 Gbit/s. When the date rate increases further, the eye-diagram becomes close. This can be improved slightly by increasing responsivity with a higher bias voltage because the signal-to-noise ratio is improved. For example, an improved eye-diagram even with a bit rate of 4.0 Gbit/s is achieved by increasing the bias voltage to 2.5V, as shown in Fig. 6(e).

## PMMA cladding's influence on the devices

In order to protect BP from the degradation due to the oxidation in air, a potential approach is introducing some specific protection layer as the upper-cladding. Here, we introducing a very convenient way of using a 500nm-thick PMMA spin-coated on the top of chip. From the measured frequency responses for the photodetectors with and without the PMMA upper-cladding (shown in Fig. S8 in Supplementary), it can be seen that the PMMA upper-cladding has little influence on the 3 dB-bandwidth as well as the responsivity of the photodetector. We observed that the resistance of the photodetectors did not change notably and the photodetectors still work well with high speed and high responsivity even when the device is 6-month old. This experiment verified that the PMMA upper-cladding greatly helps the protection of the BP film.

## Conclusion

We have demonstrated a high-performance hybrid silicon/BP waveguide photodetector for the 2 μm wavelength-band. For the fabricated photodetector, the responsivity is up to 306.7 mA/W even with a low bias voltage of 0.4 V. Our experimental results have shown that the responsivity is strongly dependent on the BP orientation. It is helpful to maximize the responsivity when the BP film is aligned so that the armchair axis become perpendicular to the dominant electric-component $E_x$ of the TE$_0$ mode. For the present photodetector, the measured 3dB-bandwidth is as large as 1.33 GHz. This high-speed and high-responsivity hybrid silicon/BP waveguide photodetector has further been used to realize a 4.0 Gbit/s data transmission at 2 μm. The impedance characterization and the established equivalent circuit model have shown that the response speed of the present silicon/BP waveguide photodetector is mainly limited by the $RC$-constant. It is possible to further improve the response speed by reducing the BP-metal contact resistance as well as the BP capacitance. The robustness of the present photodetectors has also been characterized. It has also been shown that the silicon/BP waveguide photodetector is robust when a PMMA upper-cladding was introduced as the protection layer. Our experimental results have also shown that the present photodetector is not sensitive to the environmental temperature. The operation wavelength of the silicon/BP waveguide photodetector can also be extended to be longer regarding the narrow energy bandgap of BP[42]. This is helpful for developing mid-IR silicon photonic integrated circuits in the future.

## Methods

The hybrid silicon/BP waveguide photodetectors were fabricated on an SOI wafer with a 340 nm-thick top-silicon layer. The structures were formed by using the E-beam lithography followed by the inductively-coupled-plasma (ICP) etching process with the mixed gases of $C_4F_8$ and $SF_6$. The etching depth for the SOI ridge waveguides and the grating couplers is 220nm. Then a 20nm-thick $Al_2O_3$ layer was deposited on the top of the chip to serve as the insulator layer between silicon core and the BP layer. Finally, the metal pads (15 nm Ti/80 nm Au) were formed by using the electron-beam evaporation process and the lift-off process.

The BP flake was exfoliated by using the tapes and was then transferred to the surface of a PDMS slice. Then the PDMS slice was contacted tightly to a glass sheet. It is possible to choose the piece of BP flake appropriately with the help of an optical microscopy. The chosen BP flake was transferred to cover the silicon core and the metal pads by using precise alignment.

A sourcemeter (Keithley 2400) was used for the measurement of I-V curves under the dark and illumination cases when the bias voltage varies from −0.4 V to 0.4 V. The photocurrent $I_{ph}$ was calculated from $I_{ph} = I_{ON} - I_{OFF}$, where $I_{ON}$ is the total current under illumination, and $I_{OFF}$ is the dark current. A $Tm^{2+}$-doped fiber laser at 1.95 μm was used as the light source. A variable optical attenuator (VOA) and a power splitter were used to adjust the input optical power. For the power splitter, one of the output ports was used for power monition and the other one is connected with the polarization controller. The polarization controller was inserted before the electro-optic modulator @ 1.95 μm. The RF signal was injected from a vector network analyzer (VNA, Rohde & Schwarz ZVA40, Bandwidth: 10 MHz to 40 GHz). The modulated light was coupled to the waveguide photodetector by using grating couplers. The AC photocurrent generated in the photodetector was then extracted by using a Bias-Tee (CONNPHY, Bandwidth: 30 kHz ~ 40 GHz). An electrical amplifier (OA4MVM2, bandwidth up to 30 GHz) was used to amplify the extracted AC photocurrent. Finally, the amplified electrical signal was then transmitted into the VNA and the frequency response was measured. For eye diagram measurement, the modulator RF source was from the Signal Quality Analyzer (Anritsu, MT1810A: up to 12.5 Gbit/s). And the amplified electrical signal was transmitted into the Wideband Oscilloscope (Agilent 86100C, Bandwidth: 50 MHz ~ 40 GHz). For the total impedance, the $S_{11}$ parameter could be obtained from the VNA measurement.

**Acknowledgment**
This project is supported by National Science Fund for Distinguished Young Scholars (61725503), Zhejiang Provincial Natural Science Foundation (Z18F050002); National Natural Science Foundation of China (NSFC) (61431166001, 1171101320, 61435010); and National Major Research and Development Program (No. 2016YFB0402502).

**Author contributions**
Y. Y. and J. L. prepared the fabrication of the chip. R. C., H. W., and Y. Y. preformed the BP transfer process. Y. Y. and C. L. performed the DC measurements and the frequency responses. X. F. and Y. Y. performed the eye-diagram measurement. W. D. and Y. Y. examined the BP films by using AFM. A. Q. and Y. Y. did the Raman spectrum measurement. Y. Y., J. G., and D. D. performed theoretical modeling, and data analyses. Y. Y., J. G., and D. D. wrote the manuscript. Y. Y., R. C., J. G., C. L., J. L., X. F., H. W., W. D., A. Q., S. G., H. Z., Y. S., L. T., Y. X., and D. D. contributed to the discussions and the manuscript revisions. D. D. supervised the project.

**Additional information**
Supplementary information is available in the online version of the paper. Correspondence and requests for materials should be addressed to D. D.

**Competing financial interests**
The authors declare no competing financial interests.

# Supplementary Information

# High-speed and high-responsivity hybrid silicon/black-phosphorus waveguide photodetectors at 2μm

Yanlong Yin[1], Rui Cao[2], Jingshu Guo[1], Chaoyue Liu[1], Jiang Li[1], Xianglian Feng[1], Huide Wang[2], Wei Du[1], Akeel Qadir[3], Han Zhang[2], Yungui Ma[1], Shiming Gao[1], Yang Xu[3], Yaocheng Shi[1], Limin Tong[1], and Daoxin Dai[1,†]

[1]*State Key Laboratory for Modern Optical Instrumentation, College of Optical Science and Engineering, Zhejiang University, Zijingang Campus, Hangzhou, China.*

[2]*Shenzhen Engineering Laboratory of Phosphorene and Optoelectronics Collaborative Innovation Center for Optoelectronic Science and Technology and Key Laboratory of Optoelectronic Devices and Systems of Ministry of Education and Guangdong Province Shenzhen University Shenzhen 518060, P. R. China.*

[3]*College of Information Science and Electronic Engineering, Zhejiang University, Hangzhou, Zhejiang, 310027 China.*

# Supplementary Information

**S1. Summary of the reported 2D-material photodetectors.**

**S2. Fabrication Process of hybrid silicon/BP waveguide photodetectors**

**S3. Raman spectrum of the BP films on silicon ridge waveguides**

**S4. Theory for calculating the light absorption of BP.**

**S5. Measurement of light absorption of the transferred BP films on silicon**

**S6. Analysis of the influence of the carrier mobility on the response speed.**

**S7. Equivalent circuit modeling of the present hybrid silicon/BP waveguide photodetector.**

**S8. The influence of the PMMA cladding on the performance of the hybrid silicon/BP waveguide photodetectors.**

## S1. Summary of the reported 2D-material photodetectors.

Hybrid silicon/2D-material waveguide photodetectors have become very attractive for infrared photodetection. Tables S1 shows a summary for the reported photodetectors with 2D materials. Most of them were demonstrated for the wavelength-band of 1.55 μm. For example, a hybrid silicon/graphene waveguide photodetector at 1.55 μm was realized with a responsivity of 30-50 mA/W and a 3dB-bandwidth of 18 GHz by introducing asymmetric metal structures 1. Another hybrid silicon/graphene waveguide photodetector at 1.55 μm was realized with a responsivity of 0.1 A/W and a 3dB-bandwidth of 20 GHz by using a metal-doped graphene junction2. Later, very fast hybrid silicon/graphene waveguide photodetectors at 1.55 μm were demonstrated with a bandwidth of 41 GHz 3, 65 GHz 4, and >76 GHz 5. Unfortunately, their responsivities were still relatively low, e.g., 7~30 mA/W. More recently, an excellent hybrid silicon/graphene waveguide photodetector at 1.55 μm was demonstrated with high responsivity of 0.36 A/W and high 3 dB-bandwidth of 42 GHz with the help of hexagonal boron nitride and metal-graphene edge contact6. In addition to graphene, black-phosphorus (BP) is also becoming very attractive for the infrared photodetection on silicon. A silicon/BP waveguide photovoltaic photodetector at 1.55 μm was realized with a responsivity of 0.135 A/W and a 3dB-bandwidth of ~3 GHz when the BP thickness is ~11.5nm 7. Another silicon/BP waveguide photodetector at 1.55 μm was realized with a responsivity of up to 10 A/W and a 3dB-bandwidth of ~150 MHz by utilizing the photoconductive effect 8.

Table S1. Summary of the reported 2D-material photodetectors.

| λ | Material | Optical structure | Effect | Layer structure | Responsivity | Speed | Data Links |
|---|---|---|---|---|---|---|---|
| 1.3-1.6μm | Graphene1 | Waveguide | - | M/G-M-G/M | 30-50mA/W | 18GHz | - |
| | Graphene2 | Waveguide | PV | M-G Schottky | 0.1A/W@1V | 20GHz | 12.5Gbit/s |
| | Graphene3 | Waveguide | - | M/G/M | 2-16mA/W | 41GHz | 50Gbit/s |
| | Graphene4 | Waveguide | PTE | G(P)-G(N) | 76mA/W@0.3V | 65GHz | - |
| | Graphene6 | Waveguide | PTE | M/hBN-G-hBN/M | 0.36A/W@1.2V | 42GHz | - |
| | BP7 | Waveguide | PV | M/BP/M | 0.13-0.18A/W @0.4V | 3GHz | 3Gbit/s |
| | BP8 | Waveguide | PC | M/BP/M | 10A/W@1.5V | 150MHz | - |
| 2-3.9μm | Graphene9 | Waveguide | PC | Si-G hetero-junction | 0.6A/W@-1.5V | - | - |
| | BP13 | Normal-incident | PV, PG | M/BP/M | 21-47mA/W @0.2V | - | - |
| | BP14 | Normal-incident | PG | M/BP/M | 82A/W@0.5V | kHz | - |
| | BP15 | Normal-incident | PC | M/BP/M | 0.518A/W@1.2V | - | - |
| | BP (this works) | Waveguide | PV | M/BP/M | 0.14-0.30A/W @0.4V | 0.5-1.33GHz | 4Gbit/s |

PV: Photovolatic; PTE: Photothermal; PG: Photogating; PC: Photoconductive. All data in the table are from experiments.

It is noticed that these silicon/2D-material waveguide photodetectors were developed for the wavelength-band of 1.55 μm. Since graphene has a zero bandgap9 and BP has a small bandgap of 0.3 eV~1.0 eV1011, they are available potentially for the photodetection at long wavelengths (e.g., >2 μm). In 2013, a hybrid silicon/graphene waveguide photodetector at 2.75 μm was demonstrated with 0.6 A/W at −1.5 V bias voltage12. Unfortunately, high-speed photodetection has not been reported yet. Very recently, BP has also been introduced as an alternative option for mid-IR wavelengths and several normal-incident photodetectors working at 2-3.7 μm were reported13-15. For the normal-incident BP photodetector demonstrated in 201713, the achieved responsivity is 21-47 mA/W while the high-speed operation was not reported yet. In ref. 14, a normal-incident BP photodetector at 3.9 μm was reported with a responsivity of ~82 A/W when operating at kHz-modulation frequencies. More recently, another normal-incident photodetector based on a hexagonal boron nitride (hBN)/BP/hBN-sandwiched heterostructure was demonstrated with a peak extrinsic photo-responsivity of 518 mA/W at 3.4 μm when operating at very low temperature of 77 K, while the estimated operation bandwidth is about 1.3 GHz *theoretically*[15].

It can be seen that no results have been reported for high-speed and high-responsivity silicon/BP waveguide photodetectors at the wavelength of 2 μm or longer, which play an important role in mid-IR silicon photonics for many applications[16]. In this paper, we demonstrate hybrid silicon/BP waveguide photodetectors at 2 μm with a responsivity as high as ~306.7 mA/W at a low bias voltage of 0.4 V. The achieved 3 dB-bandwidth is as high as 1.33 GHz. Finally, a 4.0 Gbit/s data-receiving test is also demonstrated *experimentally*.

## S2. Fabrication Process of hybrid silicon/BP waveguide photodetectors

Figure S2 shows the fabrication process of the hybrid Silicon/BP waveguide photodetectors. Here an SOI wafer with a 340 nm-thick top-silicon layer was used. The waveguide structures were formed by using the E-beam lithography followed by the inductively-coupled-plasma (ICP) etching process with the mixed gases of $C_4F_8$ and $SF_6$. The etching depth for the SOI ridge waveguides and the grating couplers is 220nm. Then a 20nm-thick $Al_2O_3$ layer was deposited on the top of the chip to serve as the insulator layer between silicon core and the BP layer. Finally, the metal pads (15 nm Ti/80 nm Au) were formed by using the electron-beam evaporation process and the lift-off process. The BP thin film was then transferred on the top of the SOI ridge waveguide to cover the silicon ridge as well as the electrodes by using the imprint-transfer process. Finally, in order to protect the BP thin film from oxidization, a 500nm-thick PMMA upper-cladding was spin-coated on the chip.

The imprint-transfer process is described as follows. First the BP flake was exfoliated from the bulk BP by the scotch tape. The BP flake was then transferred from the tape to a transparent and flexible PDMS substrate. With the help of a microscopy, one can select the BP flakes with the desired thicknesses and sizes by checking the color of the BP thin films. The selected BP thin film was then transferred to the target area on the silicon chip by using an imprint process. In this way, the BP thin film can be bended to tightly contact with the top-surface and the sidewalls of the silicon ridge waveguide as well as the metal electrode. This helps enhance the light absorption of BP (seen Section S5).

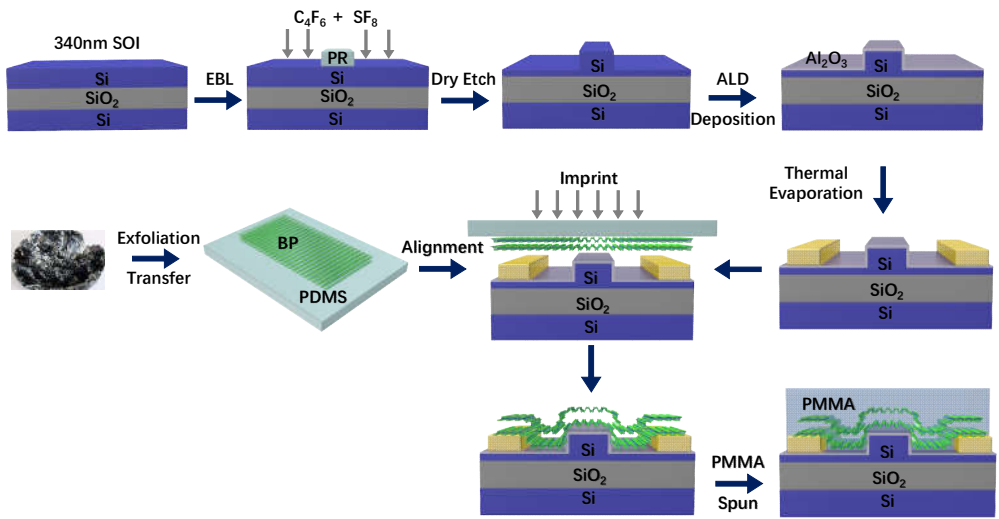

**Figure S2:** Fabrication process of the present hybrid silicon/BP waveguide photodetectors.

## S3. Raman spectrum of the BP films on silicon ridge waveguides

In order to determine the orientation angles $\theta$ of the transferred BP thin films on silicon, the samples were characterized by the Raman spectrum 717. Figure S3 (a) shows the BP structure and the two orientation axes, i.e., the armchair (AC) axis and the zigzag axis. In this experiment, the polarization angle $\theta_{pol}$ is defined as the angle between the light polarization direction and the *x* axis in the *x-y* plane, as shown in Figure S3 (a).

Figure S3 (b)-(d) shows the measured Raman spectrum for samples $S_1$, $S_2$ and $S_3$ in the cases with $\theta_{pol}$=0 and 90°. It can be seen that there are three peaks ($A^1_g$, $B_{2g}$, and $A^2_g$) as usual. The peak positions do not change almost as the polarization angle $\theta_{pol}$ varies, which is similar to that observed in ref. 13. Here the spectra are normalized to the out-of-plane $A_{1g}$ peaks since they should be independent of in-plane polarization 17. Then the orientation angle $\theta$ of the transferred BP film on silicon can be determined according to the ratio $\gamma(\theta_{pol})$ between the intensities of $B_{2g}$ peak and $A^2_g$ peak 717.

For sample $S_1$, one has $\gamma(\theta_{pol}=0)<\gamma(\theta_{pol}=90°)$, which indicates the polarization direction is close to the armchair axis when $\theta_{pol}$=90°. It means that the orientation angle $\theta$ of the BP is $\theta\approx90°$. In this case, the BP absorption of light is small when the $TE_0$ mode propagates along the hybrid silicon/BP optical waveguide. This is consistent with the measured light absorption given in Section 2 below. It is also consistent with the measured high responsivity given in the main text.

In contrast, for sample $S_3$, one has $\gamma(\theta_{pol}=0)>\gamma(\theta_{pol}=90°)$, which indicates the polarization direction is close to the armchair axis when $\theta_{pol}$=0°. It means that the orientation angle $\theta$ of the BP is $\theta\approx0°$. In this case, the BP absorption of light is high when the $TE_0$ mode propagates along the hybrid silicon/BP optical waveguide. This is consistent with the measured light absorption given in Section 2 below. It is also consistent with the measured high responsivity given in the main text.

For sample $S_2$, one has $\gamma(\theta_{pol}=0)\approx\gamma(\theta_{pol}=90°)$, which indicates the polarization direction is close to the armchair axis when $\theta_{pol}$=45°. It means that the orientation angle $\theta$ of the BP is $\theta\approx45°$. In this case, the BP absorption of light is medium when the $TE_0$ mode propagates along the hybrid silicon/BP optical waveguide. This is consistent with the measured high responsivity given in the main text.

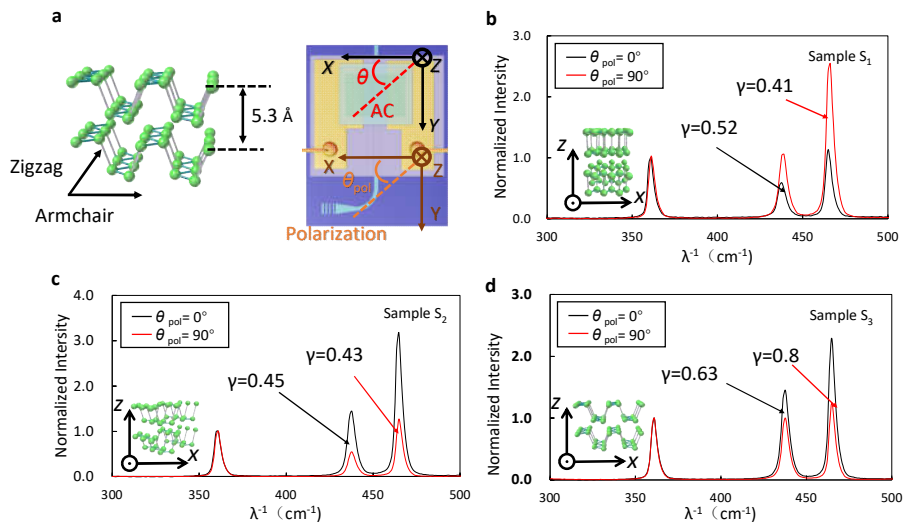

**Figure S3: a,** The BP structure and the two orientation axes (i.e., the armchair axis and the zigzag axis); **b-d,** Measured Raman spectrum for sample $S_1$ (b), $S_2$ (c), $S_3$ (d).

## S4. Theory for calculating the light absorption of BP.

The BP film with a thickness of tens of nanometers should be treated as a kind of material with anisotropic permittivity. In this calculation, we set $\theta=0°$. In this case, the $x$, $y$, and $z$ axes correspond to the armchair, zigzag, and out-of-plane axes, respectively. For the N-layer BP, the full Hamiltonian is described as 18

$$\mathbf{H} = \mathbf{H}_0 \otimes \mathbf{I}_{N \times N} + \mathbf{H}_z, \tag{1}$$

where $\mathbf{H}_z$ denotes the interlayer coupling, and $\mathbf{H}_0$ denotes the low-energy in-plane Hamiltonian around the $\Gamma$ point.

$\mathbf{H}_0$ is given as

$$\mathbf{H}_0 = \begin{bmatrix} \eta^c k_x^2 + v^c k_y^2 & \gamma k_x + \beta k_y^2 \\ \gamma k_x + \beta k_y^2 & -\eta^v k_x^2 - v^v k_y^2 \end{bmatrix}. \tag{2}$$

Here the parameter $\eta(v)^{c(v)}$ are related to in-plane effective masses of monolayer BP, $\gamma= 2.84$ eV· Å and $\beta=1.008$ eV· Å² 19 describing the effective couplings between the conduction and valence bands.

$\mathbf{H}_z$ denoting the interlayer coupling is given by

$$\mathbf{H}_z = \begin{bmatrix} \mathbf{H}^c & 0 \\ 0 & \mathbf{H}^v \end{bmatrix}, \tag{3}$$

where $\mathbf{H}^c$ and $\mathbf{H}^v$ are given by

$$\mathbf{H}^c = \begin{bmatrix} \Delta/2 & \gamma^c & & & \\ \gamma^c & \Delta/2 & \dots & & \\ & \dots & \dots & \dots & \\ & & \dots & \Delta/2 & \gamma^c \\ & & & \gamma^c & \Delta/2 \end{bmatrix},$$

$$\mathbf{H}^v = \begin{bmatrix} -\Delta/2 & \gamma^v & & & \\ \gamma^v & -\Delta/2 & \dots & & \\ & \dots & \dots & \dots & \\ & & \dots & -\Delta/2 & \gamma^v \\ & & & \gamma^v & -\Delta/2 \end{bmatrix}, \tag{4}$$

where $\Delta$ is the bandgap of the monolayer BP (i.e., $\Delta=2.12$ eV), $\gamma_c = 0.59$ eV and $\gamma_v = -0.29$ eV18. The Hamiltonian in (S1.1) is diagonalized to evaluate the eigenvalues $E_{sjk}$ and eigenvectors $\Psi_{sjk}$, in which $s$ ($c$ or $v$) denotes conduction or valence bands, $i$ and $j$ are the sub-band indices. The conductivity tensor can be written as the function of the frequency $\omega$ and $\mathbf{q}$ according to the Kubo formula, i.e.,

$$\sigma_{\alpha\beta}(\mathbf{q},\omega) = \begin{bmatrix} \sigma_{xx} & \sigma_{xy} \\ \sigma_{yx} & \sigma_{yy} \end{bmatrix}, \tag{5}$$

$$\sigma_{\alpha\beta}(\mathbf{q},\omega) = \frac{g_s \hbar e^2}{i(2\pi)^2} \sum_{ij} \int d\mathbf{k} \frac{[f(E_{cik}) - f(E_{vjk})] |\langle \Psi_{cik} | \hat{v}_\alpha | \Psi_{vjk} \rangle \langle \Psi_{vjk} | \hat{v}_\beta | \Psi_{cik} \rangle}{(E_{cik} - E_{vjk})(E_{cik} - E_{vjk} + \hbar\omega + i\eta)} \quad (\alpha, \beta=\text{x,y}), \tag{6}$$

where $g_s=2$ is the spin degeneracy, $f$ is the Fermi-Dirac distribution function in which temperature is set to 300K, $\hat{v}_{x/y} = \hbar^{-1} \partial \mathbf{H} / \partial k_{x/y}$ is the velocity operator along the $x/y$-axis, and the finite damping $\eta$ is set to 10 meV. For the local conductivity (i.e., q→0), only the diagonal elements of the tensor are non-zero. The elements of BP permittivity tensor

can be evaluated by the conductivity tensor according to

$$\varepsilon_\alpha = \varepsilon_{\alpha r} + i\frac{\sigma_{\alpha\alpha}}{\varepsilon_0 \omega h} \quad (\alpha = x, y),$$

$$\varepsilon_z = \varepsilon_{zr}, \tag{7}$$

where, $\varepsilon_{xr}$, $\varepsilon_{yr}$, and $\varepsilon_{zr}$ are the background relative permittivity tensor elements, and $h_{BP}$ is the BP thickness. In practice, $\varepsilon_{xr}$, $\varepsilon_{yr}$, and $\varepsilon_{zr}$ are taken from the measured data for bulk BP 20, and only the calculated real parts of $\sigma_{\alpha\alpha}$ are taken into account in Eq. (7) to evaluate the imaginary parts of $\varepsilon_\alpha$.

With the calculated BP permittivity tensor, the mode properties of the hybrid silicon/BP waveguide were further modeled by FEM mode-solver (from COMSOL). It should be noticed that the part of BP covering the sidewalls of the waveguide was set to be with a rotated permittivity tensor. Various values of $h$ and $\theta$ were considered for the calculation of the mode field distributions and the mode attenuation coefficients. The calculated mode attenuation coefficients are shown in Figure. 2(a) in the main text. The electric field profiles of the TE fundamental mode for the cases of $h_{bp}$=(20, 70)nm and $\theta$=(0, 90)° are shown in Figure. S4(a)-(d) as examples. It can be seen that the mode field profile has stronger field enhancement in the $Al_2O_3$ nano-slots when the BP film becomes thicker while the angle $\theta$ does not introduce significant influences on the mode field profile.

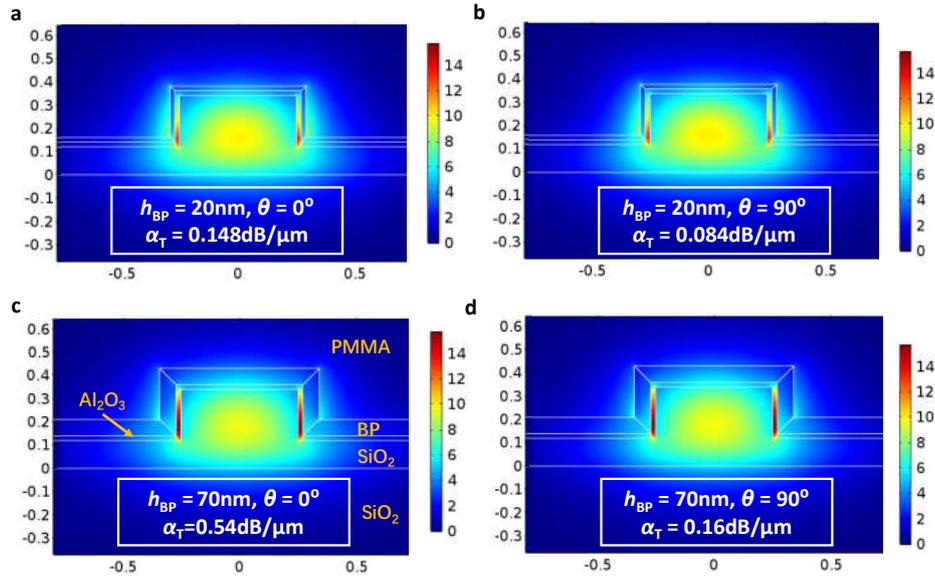

**Figure S4:** Electric field distribution |E| of the TE fundamental mode. (a) h=20 nm, and $\theta$=0° (b) h=20 nm, and $\theta$=90° (c) h=70 nm, and $\theta$=0° (d) h=70 nm, and $\theta$=90°.

## S5. Measurement of light absorption of the transferred BP films on silicon

In order to characterize the absorption of the BP film transferred on the silicon waveguide, we measured the transmissions for the silicon waveguides with/without BP films on the same chip (i.e., samples $S_1$ and $S_3$), as shown in Figure. S5(a)-(b). In our experiment, an amplifier spontaneous emission (ASE) light source with a wavelength range from 1.8μm to 2.1μm was used. From Figure. S5(c), it can be seen that the BP absorption of sample $S_3$ is much higher than that of sample $S_1$. The difference is partially caused by the difference of the BP thicknesses. More importantly, the BP orientation angle for sample $S_1$ is around $\theta=90°$ while the BP orientation angle for sample $S_3$ is around $\theta=0°$. As a result, these two samples have significantly different orientation angles $\theta$.

Here, we further calculated the BP light absorption coefficients for these three samples for different wavelengths ranging from 1.5μm to 2.5μm when the BP thin film is bended or suspended, as shown in Figure. S5(d). It can be seen that the light absorption is enhanced greatly when the BP thin film is bent. The BP light absorption is enhanced by 150%, 55%, 33% for samples S1, S2 and $S_3$, respectively. The absorption coefficient increases with the wavelength, which is due to the enhanced evanescent field. It can be predicted that the hybrid silicon/BP waveguide photodetector can even work well for long wavelength.

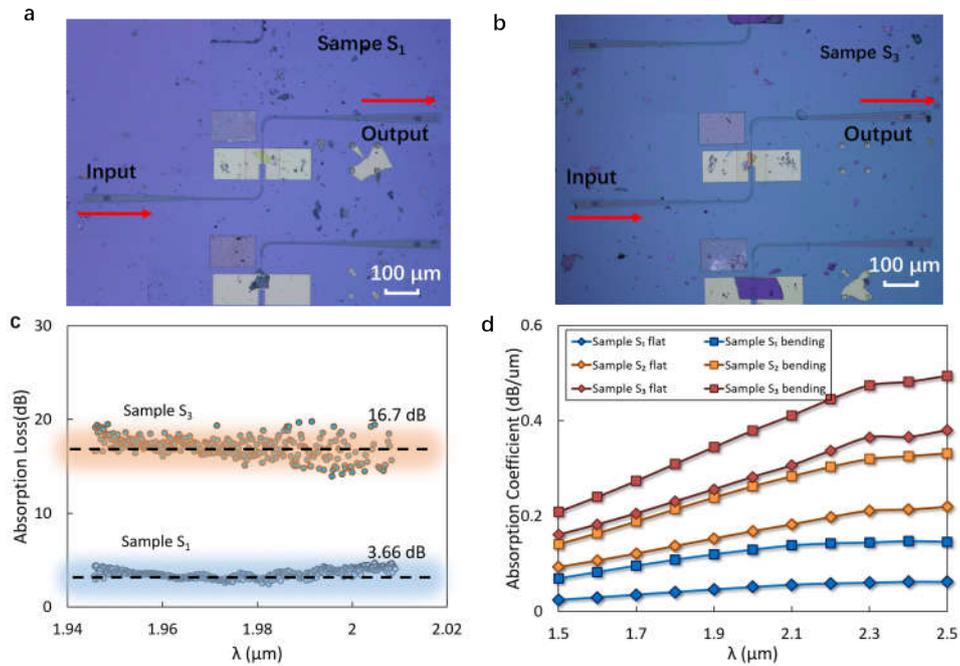

**Figure S5: a-b,** Microscopic images of sample $S_1$ and $S_3$; c, Measured absorption losses of the hybrid silicon/BP waveguide for sample $S_1$ and $S_3$; d, Calculated light absorption for the SOI ridge waveguide when the BP thin film is flat or bended.

**S6. Analysis of the influence of the carrier mobility on the response speed.**

As shown in the main text, the measured 3dB-bandwidth for the present hybrid silicon/BP waveguide photodetector is 1.33 GHz. The factors limiting the 3dB-bandwidth usually includes the transit time, and the RC-constant. In order to determine which is the dominant factor limiting the 3dB-bandwidth, here we characterized the frequency responses of the three samples ($S_1$, $S_2$, $S_3$) when operating at different bias voltages, as shown in Figure. S6 (a)-(c). It can be seen that higher responsivity was achieved with a higher bias voltage as expected while the 3dB-bandwidth of the frequency response is insensitive to the bias voltage. This indicates the carriers transport with the saturated velocity. According to a simple calculation, the transit time $\tau_{tr} = l^2/(2\mu V_{bias})$, where the channel length $l$ is 4.54μm, the $V_{bias}$ is set to be 1V. The carrier mobility is reported from $10^2 \sim 10^4 (cm^2/V \cdot s)$[17,21], and here we assume it to be $1000(cm^2/V \cdot s)$. So the transit time is about 103 ps. Based on the extracted parameters of the saturation time $\tau_s$ (64 ps) and the recombination time $\tau_r$ (240 ps) from the intrinsic BP [22], the response time $\tau$ is about 98.4 ps. The 3dB response bandwidth $f_{3dB} = 0.55/\tau \approx 5.6$ GHz. Therefore, the 3dB-bandwidth for the present hybrid silicon/BP waveguide photodetector is not limited by the transit time, and one should reduce RC constant to improve the response speed in the future.

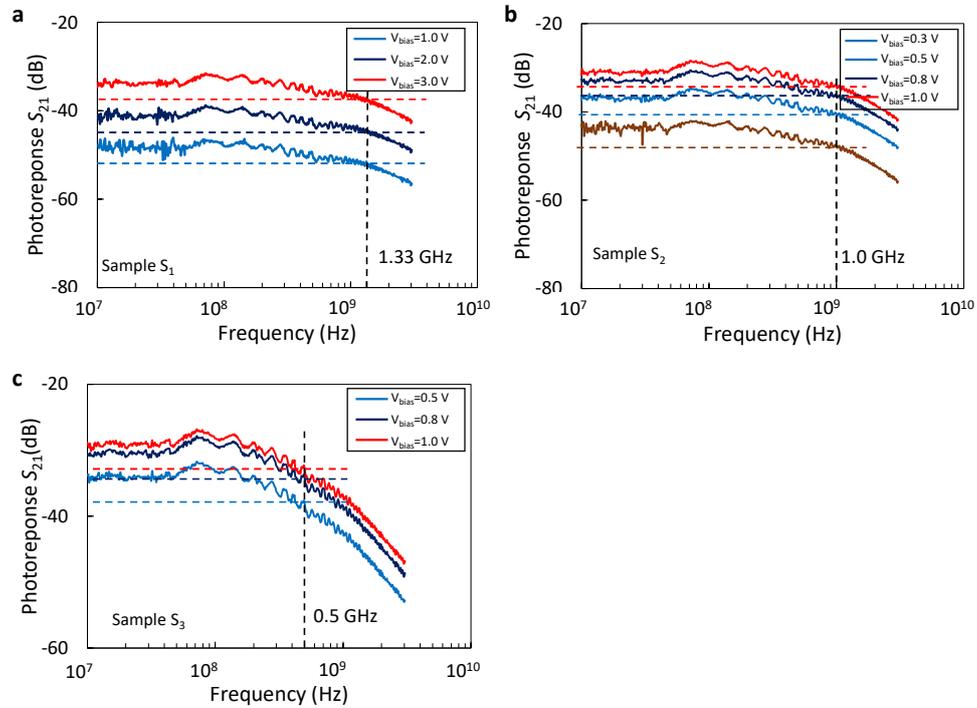

**Figure S6:** Measured frequency responses with different bias voltages for samples $S_1$ (a), $S_2$ (b), $S_3$ (c).

**S7. Equivalent circuit modeling of the present hybrid silicon/BP waveguide photodetector.**

In order to understand the RC-constant limitation to the bandwidth of the photodetector, we established the equivalent circuit model [see Figure S7(a)], in which $C_{BP}$ is the BP capacitance, $C_P$ is for the pad capacitance, $R_C$ is the BP-metal contact resistance, $R_S$ is the serial resistance, $R_{BP0}$, $R_{BP1}$, and $R_{BP2}$ are the resistances for three different parts of the BP thin film. The former two ($R_{BP0}$ and $R_{BP1}$) are for the two parts of the BP thin film with light absorption, while $R_{BP0}$ is for the part covering the silicon ridge and $R_{BP1}$ is for the part covering the silicon slab. $R_{BP2}$ is for the part of the BP thin film without light absorption. The reflection coefficient $S_{11}$ of the photodetector was then measured by using a vector network analyzer (VNA). One can easily obtain the impedance $Z_{in}$ (including the real part $Z_{re}$ and the imaginary part $Z_{im}$) from the measured $S_{11}$. The parameters for all the RCL elements in the equivalent circuit are determined by fitting the data of the real part $Z_{re}$ and the imaginary part $Z_{im}$. As an example, the fitting results for samples $S_1$, $S_2$ and $S_3$ are shown in Figure S7(b)-(d). The corresponding fitting parameters ($R_C$, $R_S$, $R_{BP0}$, $R_{BP1}$, $R_{BP2}$, $C_P$ and $C_{BP}$) are listed on the Table S2.

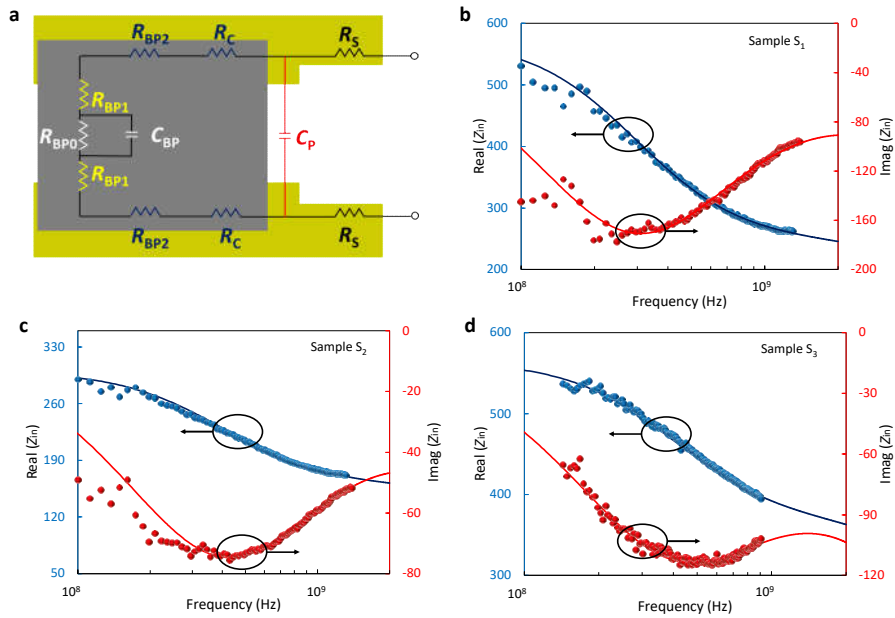

**Figure S7:** a, The established equivalent circuit of the photodetector; The fitting for the real part Real(Zin) and the imaginary part Imag(Zin) for samples $S_1$ (b), $S_2$ (c), and $S_3$ (d).

Table S2. All fitting parameters for the impedance circuit for the photodetector.

| Parameters | $R_{BP0}$ (Ω) | $R_{BP1}$ (Ω) | $R_{BP2}+R_C$ (Ω) | $C_{BP}$ (pF) | $C_P$ (fF) | $R_S$ (Ω) |
|---|---|---|---|---|---|---|
| Sample $S_1$ | 310 | 71.5 | 59 | 1.67 | 58 | 1.7 |
| Sample $S_2$ | 133.7 | 30 | 50 | 2.88 | 66 | 3.1 |
| Sample $S_3$ | 186 | 15 | 178.5 | 2.1 | 40 | 2.5 |

From Table S2, it can be seen that the pad capacitance $C_P$ and $R_S$ for these three samples are similar. Since the BP thin films for them have different orientation angles (i.e., $\theta_{S1}$=~90°, $\theta_{S2}$=~45°, $\theta_{S3}$=~0°), they have different resistances for the BP thin film. For example, the BP resistance for sample $S_3$ with $\theta_{S3}$=~0° is the smallest (see $R_{BP1}$) owning to its highest carrier mobility. The comparison of the resistance $R_{BP2}+R_C$ indicates that the contact resistance $R_C$ for sample $S_3$ is much higher than samples $S_1$ and $S_2$, which might be caused by some issues of the imprint-transfer process. From the extracted parameters ($R_C$, $C_{BP}$, $R_{BP}$), it can be predicted that sample $S_1$ has the fastest response owning to its smallest capacitance $C_{BP}$ and small contact resistance $R_{BP2}+R_C$, which is consistent with the experimental results. In the future, the response speed can be improved greatly by further reducing the BP-metal contact resistance and the BP capacitance.

**S8. The influence of the PMMA cladding on the performance of the hybrid silicon/BP waveguide photodetectors.**

BP has high risk of degradation in the air due to the surface oxidization23. In order to avoid this issue, previously samples were usually characterized within a vacuum space without oxygen and water. It is also possible to protect the sample with an $Al_2O_3$ nano-layer formed by the ALD process. This introduces some inconvenience for the measurement and some complexity for the fabrication. Here, we introduced a 500nm-thick PMMA thin film on the top of the silicon chip as for protection. Relying on the hydrophobicity, the PMMA thin film could prevent the BP thin film from the oxidation due to the water-vapor and oxygen in air. The frequency responses $S_{21}$ of the samples before and after the PMMA upper-cladding were measured, as shown in Figure. S8. It can be seen that the PMMA upper-cladding does not introduce any influence almost. The devices performance remains well when it was 6-months-old.

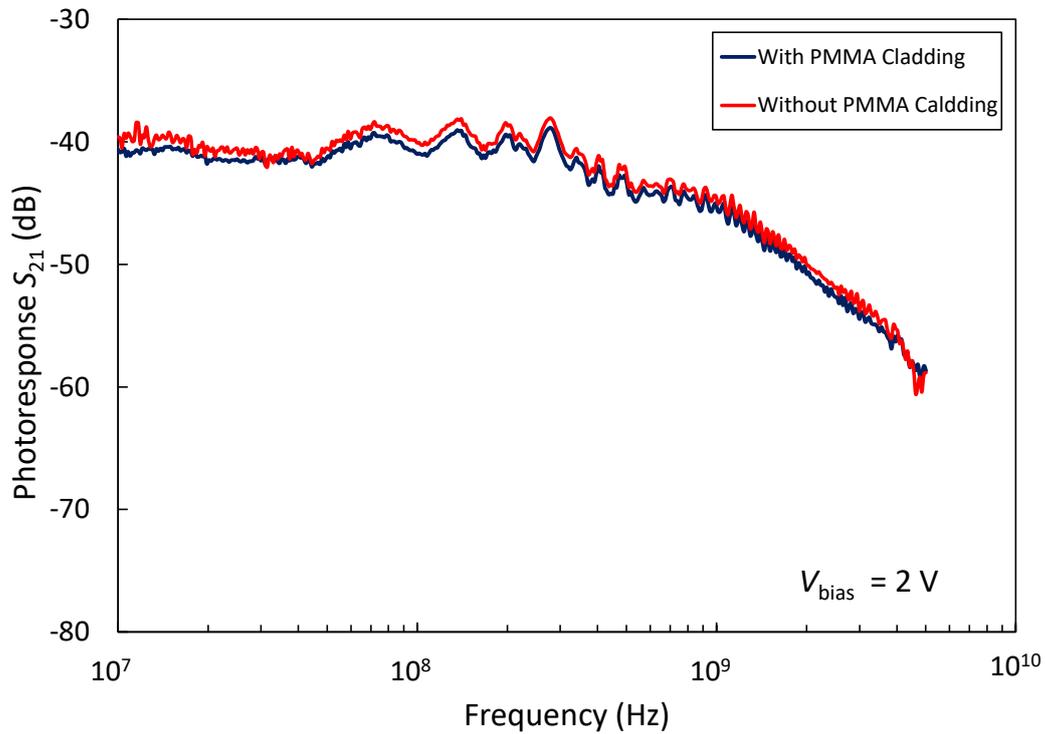

**Figure S8:** Measured frequency responses $S_{21}$ of the samples before and after the PMMA upper-cladding was introduced.

## Supplementary Reference